# Inconsistencies in the description of pairing effects in nuclear level densities

Karl-Heinz Schmidt and Beatriz Jurado

*CNRS, IN2P3, CENBG, UMR 5797, Chemin du Solarium B.P. 120, F-33175 Gradignan, France*

**Abstract:** Pairing correlations have a strong influence on nuclear level densities. Empirical descriptions and theoretical models have been developed to take these effects into account. The present article discusses cases, where descriptions of nuclear level densities are inconsistent or in conflict with the present understanding of nuclear properties. Phenomenological approaches consider a back-shift parameter. However, the absolute magnitude of the back-shift, which actually corresponds to the pairing condensation energy, is generally not compatible with the observation that stable pairing correlations are present in essentially all nuclei. It is also shown that in the BCS model pairing condensation energies and critical pairing energies are inconsistent for light nuclei. A modification to the composite Gilbert-Cameron level-density description is proposed, and the use of more realistic pairing theories is suggested.

*Introduction*

Nuclear level densities are a very important ingredient for the description of almost any nuclear reaction, either for calculating the phase space in the statistical model or for determining the driving force in dynamical models. Therefore, much effort has been invested to provide appropriate formulae or data tables, reaching from simple empirical descriptions to elaborate microscopical model calculations. Previous studies aimed to carefully adjust these descriptions to the available experimental data (e.g. ref. [1]). However, since experimental knowledge is rather restricted, mostly to energies below the neutron separation energy, only local adjustments could be achieved. In contrast, the present work aims for a deeper analysis of the different level-density descriptions, which are presently in use, over a wider excitation-energy range. It will be shown that some of these descriptions are inconsistent or in conflict with our present understanding of nuclear properties.

*Salient features of nuclear level densities*

Assuming that the nucleons move independently in the nuclear-potential well, the nuclear level density is given as the number of possible single-particle excitations in a given excitation-energy interval. The nuclear excitation energy is given by the sum of particle and hole energies with respect to the Fermi energy. The nuclear level density in the independent-particle model may be computed exactly by combinatorial methods [2, 3]. By assuming that the distance between neighbouring single-particle levels around the Fermi energy is constant, Bethe [4] derived an analytical expression for the density of nuclear states as a function of the excitation energy $E^*$, which is behind the following Fermi-gas formula:

$$\omega_F^{tot}(E^*) = \frac{\sqrt{\pi}}{12} \frac{\exp[2\sqrt{aE^*}]}{a^{1/4}(E^*)^{5/4}}. \tag{1}$$

The level-density parameter $a$ is connected with the single-particle level density $g$ by $a = g\pi^2/6$. The prominent feature of the Bethe formula is the exponential term. The pre-exponential factor mostly influences the behavior at low excitation energies. Some deficiency of the analytical formula (1) at very low excitation energies may be removed by a correction proposed by Grossjean and Feldmeier [5] using a modified saddle-point method and by Jelovic [6] with statistical techniques.

There are mainly three effects that modify the nuclear level density with respect to the Bethe formula substantially, mostly at relatively low excitation energies. First, the single-particle level density fluctuates due to shell effects, in particular at magic numbers due to large gaps in the single-particle level scheme. Secondly, the collective levels (e.g. rotations and vibrations) are not accounted for in the independent-particle model. Thirdly, the independent-particle picture is also violated by pairing correlations below the critical pairing energy. A concise description of these effects is given e.g. in ref. [7].

Nowadays, advanced models are available for calculating the nuclear level density with microscopic methods e.g. refs. [8, 2, 9], where all the effects mentioned above are included. However, there are reasons why an explicit consideration of the different global features is useful: First, it helps to analyze and to better understand the results of these most advanced models. In particular, the influences of the different effects on numerically calculated level densities may be disentangled by analyzing their characteristic features. Secondly, relatively simple analytical formulae without an explicit theoretical justification are still used very often when nuclear level densities play a role e.g. in technical applications [7]. Thus, it is important to check to which extent these formulae are consistent with theoretical concepts and ideas.

Shell effects in the single-particle level density are most directly considered in the combinatorial approach [2]. The influence on the nuclear binding, which may exceed 10 MeV, can be determined with the Strutinsky method [10]. Systematic investigations revealed that the influence of shell effects is generally washed out with increasing excitation energy by essentially an exponential damping in the level-density parameter $a$ [11]. A convenient approach to understand the influence of shell effects on the nuclear level density is to first consider a nucleus with a smooth single-particle level scheme as obtained with the Strutinsky averaging method. This fictive nucleus would have a ground-state energy as given by a macroscopic model, e.g. the Thomas-Fermi model [12] or the liquid-drop model [13], and a level density according to the Fermi-gas formula. Fluctuations in the single-particle level scheme modify the nuclear binding energy and shift also the energies of the excited nuclear levels. This shift decreases systematically for nuclear levels at higher energies. The asymptotic behavior at high excitation energy approaches the Bethe formula based on the fictive macroscopic nuclear ground state.

Collective levels are known to strongly enhance the number of excited levels per energy interval in the range of low excitation energies, which is accessible to spectroscopy. According to Bohr and Mottelson [14], it is expected that collective excitations are built up on top of the ground state and on top of each single-particle state. Rotational bands were predicted to enhance the nuclear level density by about a factor of 50 in well deformed nuclei [15]. Collective excitations that can be expressed as a coherent superposition of single-particle excitations are expected to disappear at higher excitation energies, if the nuclear temperature becomes comparable with the energies of single-particle excitations [15]. However, the energy range, where this happens, is still under debate, since the available experimental indications are rather indirect and contradictory [16, 17].

Short-range residual interactions in nuclei lead to effects similar to superconductivity in metals or super-fluidity in liquids. They induce many interesting phenomena. One of those is an increase of the nuclear binding energy. In this sense, the influence of pairing correlations on the nuclear level density resembles the one of shell effects. However, pairing correlations only exist in a region of low excitation energies and angular momentum [18]. Recent experimental results suggest that the nuclear level density in the range of strong pairing correlations or even beyond shows a nearly constant logarithmic slope, which is equivalent to a nearly constant nuclear temperature [19]. A constant nuclear temperature would imply an infinitely large specific heat. Theoretical models show a strongly enhanced specific heat due to the phase transition [20, 21, 22] which has similarities with the superfluid-normal phase transition in some liquids.

Theoretical models, e.g. the super-fluid model [23] or microscopic models [2, 8, 9] should comply with these expectations in the frame of the respective model. This is not necessarily the case for empirical parameterisations of the nuclear level density.

*Back-shifted Fermi-gas model*

The most striking effect of pairing correlations in nuclei is the even-odd staggering of the nuclear binding energies by the pairing-gap parameter $\Delta_0 \approx 12/\sqrt{A}$ [24]. In addition, it was recognized that for the first excited states this staggering is strongly reduced and gradually disappears with increasing excitation energy (see e.g. Fig. 4 in ref. [25]). This observation lead to the insertion of an energy back-shift $\Delta$ in the Fermi-gas level-density formula, see e.g. ref. [7].

$$\omega_F^{tot}(U) = \frac{\sqrt{\pi}}{12} \frac{\exp\left[2\sqrt{aU}\right]}{a^{1/4}U^{5/4}}. \qquad (2)$$

With $U = E^* - \Delta$. In the back-shifted Fermi-gas model, the level-density parameter $a$ and the back-shift $\Delta$ are adjusted to the available data, the counting of levels at low excitation energies and the resonances at the neutron separation energy.

The level-density parameter $a$ takes into account the influence of the shell effect in the energy range up to the neutron separation energy [26]. The back-shift $\Delta$ of this model denotes the displacement of the energy scale $U$ of the Fermi-gas formula from the excitation energy of the nucleus $E^*$. As mentioned before, the Fermi-gas formula has been derived in the independent-particle picture, i.e. without considering the influence of residual interactions. However, a back-shift must be applied because the nuclear excitation energy is counted from the ground state, which is strongly affected by pairing correlations. Therefore, the back-shift corresponds to the gain of binding energy in the nuclear ground state due to pairing correlations. In most formulations of the back-shifted Fermi-gas model the value of the back shift $\Delta$ is close to zero for odd-$A$ nuclei, positive for even-even and negative for odd-odd nuclei. This means that, according to the back-shifted Fermi-gas model, pairing correlations tend to increase the binding energy of even-even nuclei, have little effect on the binding energy of odd-$A$ nuclei and reduce the binding of odd-odd nuclei. A direct consequence of this consideration is that the back-shifted Fermi-gas model is not consistent with the appearance of pairing correlations in odd-odd nuclei, since pairing correlations are only stable if they lead to a gain in binding energy. This is in conflict with observations, which indicate the presence of pairing correlations in essentially all nuclei (may be with the exception of a few doubly magic nuclei), e.g. by the even-odd fluctuations in the binding energies, the reduced momenta of inertia [27] or the systematic deviation of the nuclear level density from the Fermi-gas description that is based on the independent-particle picture.

Even if the back-shifted Fermi-gas formula gives rather good descriptions for the level densities of some nuclei [28], it does not reflect the expected change in the heat capacity and thus in the slope of the level-density curve at the critical pairing energy, see the discussion on the nuclear temperature below. Moreover, the use of the Fermi-gas formula that is valid in the independent-particle picture in an energy range where pairing correlations are present is a severe inconsistency. This means that the back-shifted Fermi-gas model may well represent the nuclear level density in the limited energy range, where the parameters were adjusted, but it is expected that it fails to properly describe the level density outside this range, e. g. at higher energies [29].

*Composite Gilbert-Cameron level density*

The composite Gilbert-Cameron level-density formula [30] is composed of a constant-temperature formula below and a back-shifted Fermi-gas formula above a matching energy, where the two descriptions join continuously with identical slopes. The constant-temperature part was originally introduced to compensate for the low-energy collective levels not included in the Fermi gas model. There exist different parameterisations. In the one proposed in RIPL-3 [7], the back-shift parameter $\Delta$ of the Fermi-gas description above the matching energy is zero for odd-odd nuclei, $\Delta_0$ for odd-mass nuclei and $2\Delta_0$ for even-even nuclei. In this description, the energy gain by pairing correlations in odd-odd nuclei is zero, suggesting that there is no pairing in odd-odd nuclei. Since the first excited quasi-particle state in even-even nuclei has the same number of unpaired particles as an odd-odd nucleus in its ground state, one would also expect that pairing disappears at the first excited quasi-particle state in even-even nuclei. This is again in severe conflict with the presence of pairing correlations in essentially all nuclei, even at moderate excitation energies, with the eventual exception of a few doubly magic nuclei.

This problem might be solved by assuming a higher matching energy in the Gilbert-Cameron composite level density. However, the matching conditions would require a considerably increased level density in the independent-particle regime. Since the value of the level-density parameter $a$ is rather well established on theoretical ground, this increase would be consistent with the expected presence of a collective enhancement of the nuclear level density. Note that the collective enhancement only weakly depends on excitation energy and thus has little effect on the level-density parameter[1].

An example of the composite Gilbert-Cameron level density is shown in Fig. 1. In addition to the conventional composite Gilbert-Cameron level density, a modified description is shown. In this description, the level density in the Fermi-gas regime is increased by a factor of 50, which represents well the magnitude of the collective enhancement in deformed nuclei. The matching condition with the empirical constant-temperature part of the level density demands an increase of the back-shift of the Fermi-gas part by about 2.5 MeV.

---

[1] Compared to the variation of the level density itself, the variation of the collective enhancement is slow. E.g. the level density of a heavy nucleus increases by a factor of about $10^5$ if the excitation energy increases from 20 MeV to 30 MeV. The collective enhancement, however, can vary from some maximum value around 50 to 1, only. Theoretically, one expects that this variation occurs over a rather large energy interval of about 10 MeV ( see G. Hansen, A. S. Jensen , Nucl. Phys. A **406**, 23 (1983)).

Empirical information on the critical energy, the excitation energy where pairing correlations disappear, has been extracted from angular anisotropy in low-energy fission [31], where values around 10 MeV have been deduced. These values are indeed appreciably larger than the matching energies normally applied in the composite Gilbert-Cameron level densities [7]. It seems natural to associate the critical pairing energy with the matching energy of the Gilbert-Cameron composite level-density description.

In the light of recent experimental results, which indicate that the nuclear level density in the regime of pairing correlations is well approximated by a constant temperature, especially for heavier nuclei [19, 32], the composite Gilbert-Cameron level-density description is very appealing, provided that the back-shift is considerably increased. In detail, a modified composite level-density formula could unambiguously be constructed by describing the constant-temperature part with an empirical systematics, e.g. from ref. [26] and by borrowing the energy-dependent level-density parameter for the Fermi-gas part from ref. [11]. The composite formula is fully determined by fixing the matching energy. The matching energy is uniquely determined if there is some reliable information available on the critical pairing energy or if the magnitude of the collective enhancement in the Fermi-gas regime is imposed. The back-shift parameter of the Fermi-gas part is then given by the usual matching conditions. The fade out of the collective enhancement remains an open problem, but experimental results [16] and theoretical arguments [17] suggest that this feature is rather inconspicuous.

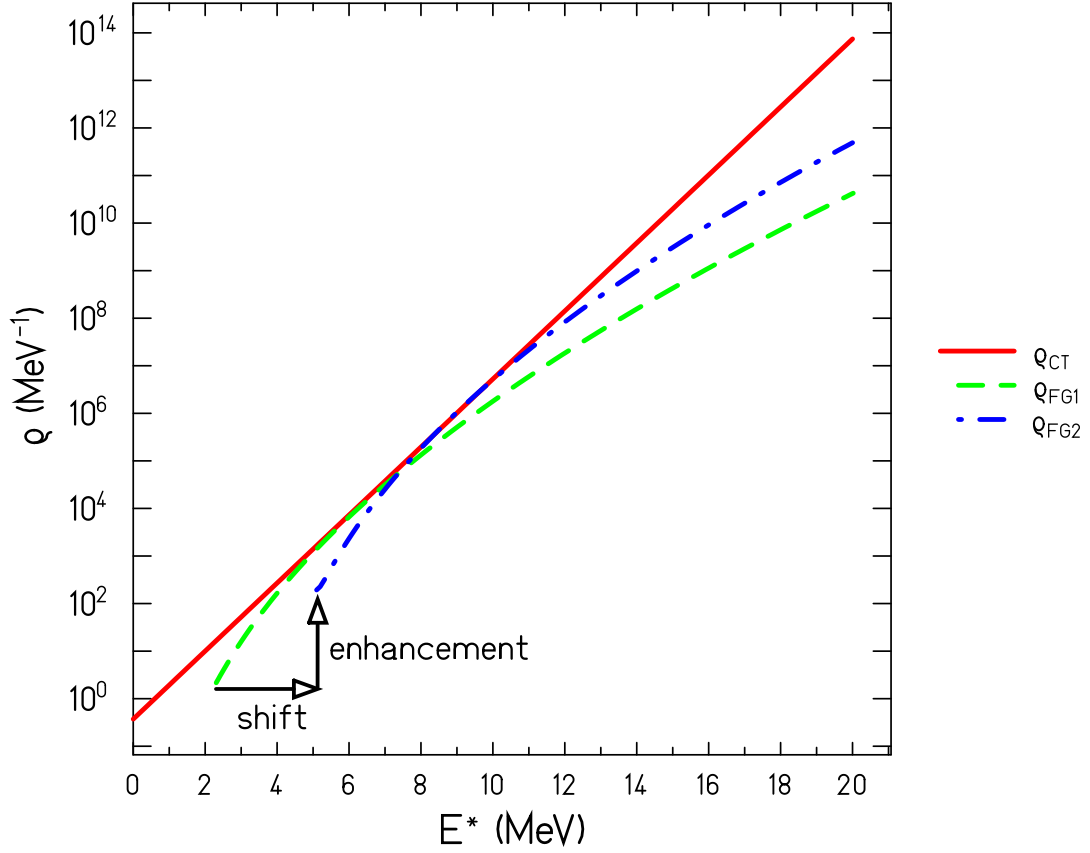

**Figure 1:** Schematic illustration of the composite Gilbert-Cameron level-density formula and a modification with increased matching energy for $^{154}$Sm. Red full line: Constant-temperature part. Green dashed line: Back-shifted Fermi-gas part of the original Gilbert-Cameron description. Blue dash-dotted line: Modified back-shifted Fermi-gas part of the composite level-density formula with increased back-shift and additional enhancement factor. The back shift is increased by about 2.5 MeV, and the level density in the Fermi-gas regime is enhanced by about a factor of 50 with respect to the conventional composite Gilbert-Cameron level density. The constant-temperature part is valid below, and the Fermi-gas part is valid above the respective matching energy.

*BCS model*

The effect of pairing correlations on the level density can be included e.g. in the partition-function method [33]. According to the BCS model [34], adapted to nuclei [35], pairing correlations lead to additional nuclear binding by the condensation energy

$$E_{cond} = \frac{1}{4}g\Delta_0^2 - n\Delta_0. \tag{3}$$

with $n=0$ for even-even, $n=1$ for odd-mass, and $n=2$ for odd-odd nuclei [7]. An equidistant single-particle level scheme with density $g$ is assumed. The condensation energy is shown in Fig. 2 as a function of mass number for odd-odd, odd-$A$ and even-even nuclei. It is assumed that the pairing-gap parameter is given by $\Delta_0 \approx 12/\sqrt{A}$ and the level-density parameter is $a = g\pi^2/6 = A/10$. A comprehensive analysis of the generalized super-fluid model [23], which is compatible with the BCS model [36], reveals a good agreement with a variety of experimental data [37]. However, there appears a problem for lighter nuclei: As shown in Fig. 2, for odd-odd nuclei, pairing correlations do not increase the nuclear binding energy for nuclei with $A < 110$. The same is true for odd-$A$ nuclei with $A < 25$. Thus, pairing correlations should not be stable in odd-$A$ and odd-odd nuclei below these limits according to the BCS model, which contradicts experimental observation.

Fig. 2 also reveals an internal inconsistency of the BCS model. In addition to the condensation energy, the critical excitation energy $U_C = aT_C^2 + E_{cond}$ (with the critical temperature $T_C = 0.567\Delta_0$) above the ground state of the respective nucleus without pairing correlations is shown by the dotted line marked by $U_c$-$E_{cond}$. For all nuclei with stable pairing correlations, nuclear levels below the critical pairing energy should shift to lower energies. Thus, the condensation energy should have positive values, and the pairing condensation energy should increase the nuclear binding energy. This is not the case. On the one hand, the turn-over of -$E_{cond}$ to positive values near $A=110$ for odd-odd nuclei and near $A=25$ for odd-$A$ nuclei indicates that pairing correlations are not stable below these respective limits, as mentioned already above. On the other hand, the critical energy reaches the ground-state energy near $A=15$ for odd-odd nuclei and near $A=4$ for odd-$A$ nuclei, suggesting that pairing correlations are stable down to these respective limits, which are appreciably lower.

Thus, there exists an internal inconsistency in the theory. Of course, the specific values determined above depend on the pairing strength and the single-particle level density, but this does not remedy the fundamental problem evidenced in this specific case, where we use well-founded values for $\Delta_0$ and $g$ [37].

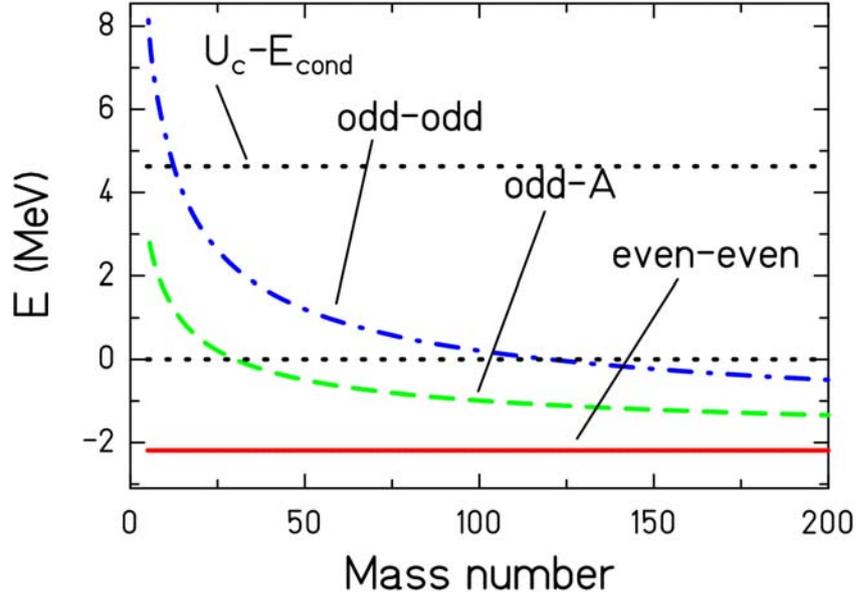

**Figure 2:** Negative BCS pairing condensation energy according to eq. (3) for even-even (full line), odd-$A$ (dashed line) and odd-odd (dash-dotted line) nuclei. Negative values correspond to the gain in nuclear binding energy by pairing correlations, positive values mean that pairing correlations would not increase the nuclear binding energy. $U_c$ denotes the critical energy, above which pairing correlations disappear according to the BCS model. The dashed line marked by $U_c$-$E_{cond}$ corresponds to $U_C - E_{cond} = aT_C^2$. See text for details.

This result sheds doubts on the applicability of the BCS theory to pairing correlations, at least in light nuclei. May be, the reason is an underestimation of the magnitude of the condensation energy in the BCS approach, which has been claimed [38, 39] by a comparison with a calculation on the basis of the Richardson approach [40]. More realistic results are expected from more general exactly solvable pairing models [41].

*Nuclear temperature*

In many aspects, the excitation energy dependence of the nuclear level density is more decisive than its absolute value. E.g. it determines the shape of the energy distributions in evaporation [42, 43], and it governs the partition of excitation energy in binary reactions in statistical equilibrium [44, 45]. Therefore, the temperature, defined as $T = (\frac{d \ln \rho}{dE})^{-1}$ is compared as a function of excitation energy in Fig. 3 for different empirical and theoretical descriptions.

The super-fluid model and several microscopic models clearly show the different behavior in the super-fluid regime and above. The same is true for the composite level-density formula. In these descriptions the nuclear temperature grows slowly or is even constant in the super-fluid regime, while it increases gradually with a decreasing slope in the Fermi-gas regime. This is not the case for the back-shifted Fermi-gas model, which reveals that this description is in conflict with the expected effect of pairing correlations on the nuclear level density below the critical pairing energy. See also the discussion on this subject in refs. [37, 46, 47]. It should also be considered that the sharp phase transition at the critical energy, e.g. supposed in the composite Gilbert-Cameron description, is not realistic, since the phase transition is washed out due to the finiteness of the nuclear system [48]. The BCS level density with the average pairing gap shown in Fig. 3 demonstrates this behavior. Furthermore, one cannot exclude the influence of other kinds of residual interactions, e.g. those which are behind the congruence energy [49]. Those might act up to higher excitation energies and shift the transition to the Fermi-gas regime to even higher energies. But at sufficiently high excitation energy, the independent-particle model is expected to be valid, and thus the constant-temperature approach [50, 7], which assumes a constant logarithmic slope of the level density at all energies, should not be valid beyond this limit. Therefore, a model based only on the constant-temperature formula is not explicitly discussed in this work.

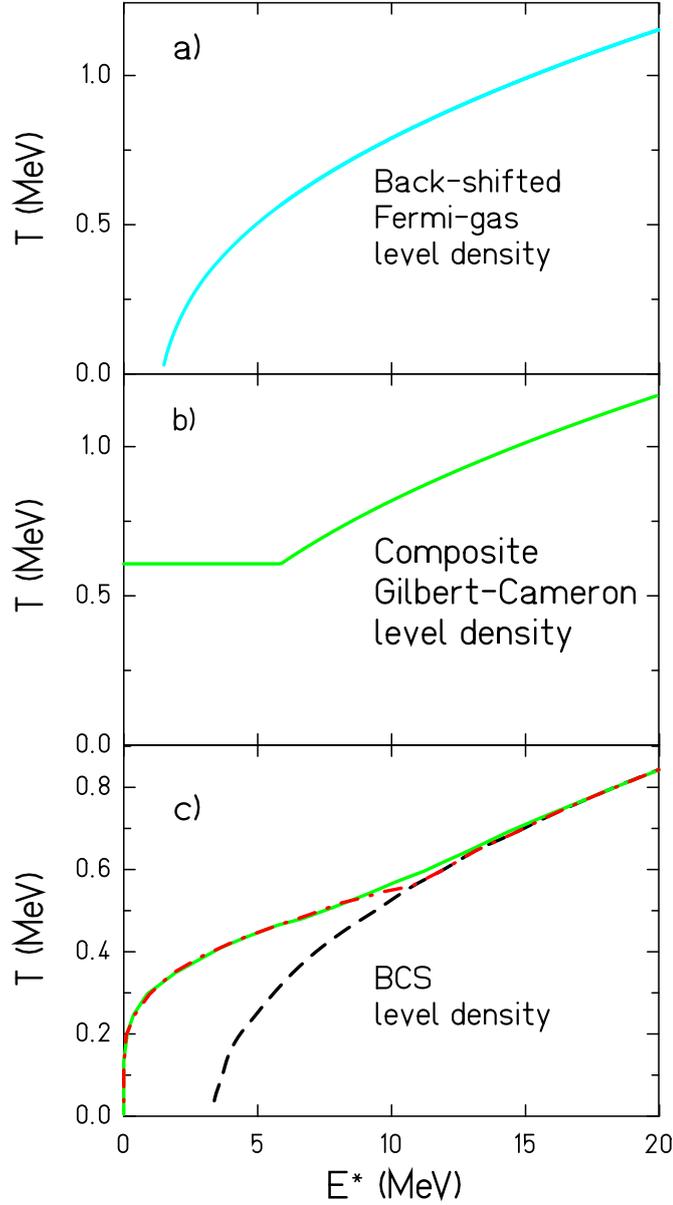

**Figure 3:** Schematic illustration of the behavior of the nuclear temperature in different level-density descriptions. The back-shifted level density [7] (a) and the composite Gilbert-Cameron formula [7] (b) for $^{154}$Sm are compared with the BCS level density (data taken from ref. [48]) for a nucleus with $g$ = 7/MeV (c). The different curves in (c) denote the Fermi-gas level density without pairing interactions extrapolated from energies far above the critical energy (dashed black line), the BCS level density with the most probable pairing gap (red dash-dotted line), and the BCS level density with the average pairing gap (green full curve).

*Conclusion*

Empirical and theoretical descriptions of the nuclear level density consider the influence of pairing correlations in different ways. In the back-shifted Fermi-gas level density formula, the even-odd staggering of the nuclear binding energy is compensated by a corresponding purely empirical staggering of the back-shift parameter obtained from a fit to experimental level densities. In the composite Gilbert-Cameron level-density description, there is an additional constant-temperature energy range, which was originally introduced to compensate for the low-energy collective levels not included in the Fermi-gas model. In effect, this low-energy part also accounts qualitatively very well for the increased heat capacity due to pairing correlations below the critical pairing energy. Microscopic models calculate the quasi-particle excitations in the energy range of pairing correlations, e.g. with the BCS pairing theory, and thus explicitly consider the deviations from the independent-particle picture, which is supposed to be valid at higher energies in the Fermi-gas regime.

Our analysis revealed that all these empirical and theoretical descriptions of the nuclear level density are incompatible with the experimentally known presence of pairing correlations in essentially all nuclei. The back-shifted Fermi-gas model does not account for the expected deviation of the nuclear properties from the independent-particle picture, in particular an increased heat capacity, in the super-fluid regime. Also, the back-shifted Fermi-gas description and the composite Gilbert-Cameron formula were analyzed in view of their consistency with the expected manifestations of pairing correlations in nuclear level densities. It was found that these formulae predict no gain in energy by pairing correlations in many nuclei where pairing is known to be present. It must be concluded that these formulae are to be considered as local fits to the scarce data in a limited excitation-energy range. They are supposed to yield unrealistic results when applied beyond this range. One could argue that these descriptions should only be considered as technical tools without a deeper connection to theoretical ideas. However, when they are applied to nuclear reactions, they determine the physics of the processes and, thus, they may easily lead to wrong results and conclusions.

The same kind of analysis applied to the predictions of the BCS model yielded the surprising result that this model gives inconsistent results for the critical pairing energy and the pairing correlation energy for lighter nuclei. On the one hand, the presence of a finite critical pairing energy suggests the existence of pairing correlations for even-even, odd-*A* and odd-odd nuclei down to very light masses. On the other hand, there is no gain in binding energy by pairing correlations, and thus pairing correlations are not stable in odd-odd nuclei with $A < 110$. This finding may be explained by claims that the BCS model severely underestimates the condensation energy.

The following practical conclusions may be drawn:

1.  Pairing correlations induce a deviation of the level density from the Fermi-gas formula below the critical pairing energy. In particular, it is expected that the heat capacity is enhanced below the critical pairing energy. This requirement is not respected by the back-shifted Fermi-gas level density.
2.  Since the presence of pairing correlations in practically all nuclei is well established, the back-shift parameter of the high-energy (Fermi-gas) part of any realistic level-density description needs to be large enough for the ground state of all nuclei to be energetically below the fictive macroscopic ground state on which the Fermi-gas level density is based. This requirement is violated by the back-shifted Fermi-gas model, the commonly used formulations of the composite Gilbert-Cameron level-density formula and, for light nuclei, by the BCS model. The super-fluid model and microscopic models using the BCS theory to model pairing correlations suffer from this problem. More realistic pairing theories may solve this issue.

3. The composite Gilbert-Cameron level density with an increased back-shift and an enhancement factor accounting for collective excitations in the Fermi-gas regime seems to be in close agreement with theoretical expectations up to energies well above the critical pairing energy, if the matching energy, the temperature parameter and the level-density parameter are appropriately chosen.

**Acknowledgment**

This work was supported by the European Commission within the Seventh Framework Programme through Fission-2010-ERINDA (project no.269499).